\documentstyle[epsf]{elsart}

\newcounter{eqnval}
\newif\ifnumbysec

\def\ibm1{{\rm IBM}\rule[.9mm]{0.85mm}{0.1mm}\,1}

\begin{document}
{\small
\begin{frontmatter}
\title{On the validity of the O(18) coupling scheme in N $\sim$ Z nuclei}
\author{V.S. Lac, J.P. Elliott, J.A. Evans}
\address{Centre for Theoretical Physics, \\
The University of Sussex, Falmer, Brighton, BN1 9QH, U.K.}

\begin{abstract}
Wave-functions from the newly proposed O(18) coupling scheme in IBM-3 
are compared with those from a numerical calculation fitted to some Zn, 
Ge and Se nuclei.
\end{abstract}

\end{frontmatter}

    The interacting boson model \cite{IA87}, in which nucleon pairs  
are approximated by $s$- and $d$-bosons has enjoyed some success in 
describing nuclear systematics and, in lighter nuclei, a version 
incorporating isospin symmetry, IBM-3 \cite{EW80}, has been used. 
Each boson in IBM-3 has isospin $T=1$ giving it a total of 18 states 
so that the states of $N$ bosons belong, trivially, to the  totally   
symmetric representation $[N]$ of the group U(18).  The existence of 
any chain of  subgroups  of U(18)  which contains the group  O(3) 
$\times$ SU(2), corresponding  to  the exact symmetries of  angular  
momentum  and isospin,  provides a coupling scheme with analytic  
properties.  The most natural chain below U(18) starts with  U(6) 
$\times$ SU$_{\rm c}$(3)  which  separates  the three charge states 
$M_T=1$, 0, $-1$ of  the  boson from the six ``orbital" states of $s$ 
and $d$. The isospin group SU(2) is then a sub-group of SU$_{\rm c}$(3) 
and,  below U(6)  are  the  three alternative chains via U(5), O(6) or 
SU(3) which are familiar from the earliest days of the IBM  \cite{IA87} 
and  describe  vibrational, $\gamma$-unstable and rotational nuclei 
respectively.  It is found that the U(6) $\times$ SU$_{\rm c}$(3) chain 
has validity, with the low states in numerical IBM-3  calculations  
\cite{EES85,EELL96}  containing large amounts of the  totally symmetric  
states $[N]$ of each of the two factors SU$_{\rm c}$(3) and  U(6). For 
each value of $T=N$, $N-2$, .. 1 or 0 allowed from SU$_{\rm c}$(3) the 
set of orbital states is then identical to those in IBM-1.        

   Recently two new chains starting with the subgroup O(18) of U(18) 
have been discovered \cite{G96}. One of them proceeds via O(6) $\times$ 
SU(2) followed  by the familiar reduction from the $\gamma$-unstable  
group O(6). The other goes first to O(15) $\times$ SU$_s$(2), in which 
O(15) belongs entirely to the $d$-boson and SU$_s$(2) is the isospin of  
the $s$-bosons. Then O(15) is reduced to O(5) $\times$ SU$_d$(2) and 
finally SU(2) is reached from the product SU$_s$(2) $\times$ SU$_d$(2),  
a vector-coupling of the separate isospins of $s$- and $d$-bosons.

    Since the new O(18) classifications generally break  the  U(6) 
symmetry, which is known to be reasonably good, it is important to 
test their validity in numerical work.  In this note we report  an 
analysis, in the O(18) $\supset$ O(15) basis, of some recent numerical  
IBM-3 wave-functions \cite{EELL96}. But first we illustrate the new 
classifications in the simplest non-trivial example of $N=3$, $T=1$. 
We argue that one of the new chains, O(6) $\times$ SU(2), seems to be 
of little  interest since it is identical to the old U(6) $\times$ 
SU$_{\rm c}$(3)  classification for most of the low states. 

   In the three sections of table 1 we compare the classifications for 
the case $N=3$, $T=1$ using the old chain  U(6) $\times$ SU$_{\rm c}$(3) 
$\supset$ O(6)  $\times$ SU$_{\rm c}$(3)  and the two new chains  O(18) 
$\supset$ O(6) $\times$ SU(2) and O(18) $\supset$ O(15). (For convenience 
we refer to these three chains as A, B and C respectively.)  All  three  
chains have the O(5) group in common so it is unnecessary to specify the 
$J$-values which belong to each O(5) representation, according to familiar 
rules \cite{IA87}. (For example, (00) $J=0$, (10) $J=2$, (20) $J=2,4$, 
(30) $J=0,3,4,6$.) It is sufficient to use single numbers $\alpha$ and 
$\delta$,  as in ref.\cite{G96}, for the O(18) and O(15) labels since 
only the totally  symmetric  representations occur. In each section of 
the table there must be exactly the same set of O(5) labels 
\mbox{($\tau_1$, $\tau_2$)} but they are  grouped differently in the 
different chains.
\begin{table}
\caption{The classification of states for $N=3$, $T=1$  using  the 
three group chains A) U(6) $\times$ SU$_{\rm c}$(3) $\supset$ O(6) 
$\times$ SU$_{\rm c}$(3), B) O(18) $\supset$ O(6) $\times$ SU(2)
and \mbox{C) O(18) $\supset$ O(15)}. The superscript 2 which appears  
in chain C indicates that there are two states with the given labels, 
corresponding to different values of $T_s$ and $T_d$.}
\begin{tabular}{lclclcl}\\[-0.5cm] \hline
Chain A  && SU$_{\rm c}$(3) $\equiv$ U(6) && O(6)     &&   O(5) \\
\cline{3-7} \\[-0.7cm]
         &&  [3]                && (300)    &&  (00) (10) (20) (30) \\
         &&                     && (100)    &&  (00) (10)  \\[0.1cm]
         &&  [21]               && (210)    &&  (10) (11) (20) (21) \\
         &&                     && (100)    &&  (00) (10) \\
\hline \hline
Chain B  &&  O(18)              && O(6)     &&  O(5) \\[-0.2cm]
         && $\alpha$            &&          &&       \\
\cline{3-7} \\[-0.7cm]
         &&  3                  && (300)    && (00) (10) (20) (30) \\
         &&                     && (210)    && (10) (11) (20) (21) \\
         &&                     && (100)    && (00) (10) \\[0.1cm]
         &&  1                  && (100)    && (00) (10) \\
\hline \hline
Chain C  && O(18)               && O(15)    &&  O(5) \\[-0.2cm]
         && $\alpha$            && $\delta$ &&       \\
\cline{3-7} \\[-0.7cm]
         &&  3                  && 0        && (00) \\
         &&                     && 1        && (10)$^2$ \\
         &&                     && 2        && (00) (11) (20)$^2$ \\
         &&                     && 3        && (10) (21) (30) \\[0.1cm]
         && 1                   && 0        && (00) \\
         &&                     && 1        && (10) \\
\hline
\end{tabular}
\end{table}
 
   The states expected to be most prominent at lowest energies are 
given  in the first rows of each section.  Notice first  that  the 
chains A and B have the group O(6) in common and that,  since  the 
representations (300) and (210) occur only once, those states must 
be identical in the two chains. In other words the (300) states in 
chain  B  are identical with those in chain A and have  full  U(6) 
symmetry $[3]$ while the (210) states in chain B are identical  with 
the  mixed symmetry states $[21]$ of chain A.  In general the  (100) 
states in B will be mixtures of $[3]$ and $[21]$ but these states are 
expected to lie high in energy.  We conclude that chain B does not 
have much new interest.

   Chain C is more interesting since it has nothing in common with 
the familiar chain A beyond the O(5) label. For example, the lowest 
$0^+$ state, with $\alpha=3$, $\delta=0$, $(\tau_1, \tau_2)=(00)$ 
may be expanded in the chain A basis as
\begin{equation}
-\sqrt{\frac{3}{4}}  | s^3 [3] \rangle +
 \sqrt{\frac{5}{36}} | s d^2 [3] \rangle +
 \sqrt{\frac{1}{9}}  | s d^2 [21] \rangle
\end{equation}
which,  although containing the expected dominance of the  $s$-boson 
and  of  the  full U(6)  symmetry  $[3]$,  nevertheless  contains  a 
significant  component  of $d$-bosons and of  the  ``mixed"  symmetry 
$[21]$.  The  situation  for  the lowest $2^+$ state  is  more  complex 
because, as shown in the table, there are two states with the same 
O(18),  O(15) and O(5) labels 3, 1 and (10),  corresponding to $T_s=0$ 
and 2. Their expansions are
\begin{eqnarray}
|T_s=0 \rangle &=& \sqrt{\frac{17}{36}} |s^2 d [3] \rangle + 
                   \sqrt{\frac{17}{45}} |s^2 d [21] \rangle -
                   \sqrt{\frac{7}{68}} |d^3 [3] \rangle -
                   \sqrt{\frac{4}{85}} |d^3 [21] \rangle  \nonumber \\     
|T_s=2 \rangle &=& \sqrt{\frac{4}{9}} |s^2 d [3] \rangle - 
                   \sqrt{\frac{5}{9}} |s^2 d [21] \rangle
\end{eqnarray}

Each  of these states shows very strong mixing of the U(6)  labels 
$[3]$ and $[21]$ which would be contrary to experience for the  lowest 
$2^+$ state but a simple calculation shows that the percentage of the 
full symmetry $[3]$ can be increased up to a maximum of 97\% by suitable  
combination of the two states (2). Thus, although we would not expect 
$T_s$ to be a good quantum number, the  O(18) and  O(15) labels could 
still be good.       

    In a recent paper, ref.\cite{EELL96}, we deduced an $NT$-dependent 
IBM-3  hamiltonian from  a shell-model mapping and applied it to the Ni,  
Zn, Ge and Se isotopes in the first half of the $p_{3/2}$, $f_{5/2}$, 
$p_{1/2}$ shell. In particular we saw that the low states were dominated 
by the full-symmetry U(6)  label $[N]$ with the mixed symmetry  states 
$[N-1,1]$ coming in at about 3 MeV. It is therefore of interest to 
analyse our wave-functions in the new group chain C since the examples 
above suggest that this chain is not incompatible with a coupling scheme   
close to U(6). Table 2 shows the results of such an analysis for the low 
states of each spin for the nuclei considered. The first row of numbers 
for each nucleus gives the percentage lying in the fully symmetric U(6)  
representation $[N]$ and  is copied from table 3 of ref.\cite{EELL96}.  
The second row gives  the percentage in the favoured state of chain C,  
viz. $\alpha=N$, with the smallest $\delta$ compatible with each angular  
momentum, i.e. $\delta=0$, 1, 2, and 2 for $J=0$, 2, 3 and 4, respectively 
except for $J=3$, $T=0$ when $\delta=3$.  

\begin{table}
\caption{A  comparison of the wave-function percentages  in  the
favoured representations of U(6),  first row, and of the new chain
C, O(18) $\supset$ O(15), in the second row for each nucleus. The 
favoured representations in chain C are $\alpha=N$ and $\delta=0$, 
1, 2 (3 if $T=0$) and 2 for $J=0$, 2, 3 and 4 respectively.}
\begin{tabular}{lccccccccccccc} \\[-0.5cm] \hline
& $N$ & $T$ && $0^+_1$ & $0^+_2$ & $2^+_1$ & $2^+_2$ & $2^+_3$
& $3^+_1$ & $3^+_2$ & $4^+_1$ & $4^+_2$ & $4^+_3$ \\ \hline
\nuc{62}{Zn} & 3 & 1 && 94 & 91 & 96 & 50 & 27 & 13 & 23 & 88 & 27 & 17 
\\[-0.25cm]
             &   &   && 99 & 0  & 93 & 36 & 61 & 87 &  3 & 90 & 45 & 54 \\
\nuc{64}{Zn} & 4 & 2 && 93 & 59 & 94 & 79 & 10 & 44 & 22 & 90 & 70 &  5 
\\[-0.25cm]
             &   &   && 96 & 0  & 97 & 9  & 89 & 48 & 46 & 96 & 2  & 44 \\
\nuc{64}{Ge} & 4 & 0 && 94 & 89 & 96 & 83 & 19 & 89 &  9 & 95 & 69 & 18 
\\[-0.2cm]
             &   &   && 95 & 0  & 93 & 3  & 3  & 96 & 99 & 90 & 18 & 65 \\
\nuc{66}{Ge} & 5 & 1 && 91 & 73 & 93 & 89 & 3  & 51 & 30 & 94 & 83 &  4 
\\[-0.2cm]
             &   &   && 92 & 0  & 96 & 2  & 92 & 37 & 61 & 97 &  3 & 95 \\
\nuc{68}{Se} & 6 & 0 && 91 & 80 & 93 & 91 & 90 & 87 & 10 & 94 & 90 & 86 
\\[-0.2cm]
             &   &   && 84 & 0  & 82 & 9  & 0  & 97 & 99 & 84 & 10 & 2  \\
\hline
\end{tabular}
\end{table}

   It is immediately clear from the table that both chains give good  
descriptions  of the lowest states for $J=0$, 2 and 4 but  it  is 
important to look at the comparison in more detail because in most 
cases the numbers in the table are a total percentage summed  over 
several  independent  states with the same  representation  labels  
while, in other cases, the numbers refer to unique states. In 
\nuc{62}{Zn} for example, three of the four $0^+$ states lie within  
the $[3]$ representation of U(6) whereas the figure of 99\% refers to 
a unique state  with $\alpha=3$, $\delta=0$ in the O(18) $\supset$ 
O(15) chain. In the same nucleus there is a total of seven $2^+$ 
states of which three lie within the U(6) label $[3]$ and only two 
within $\alpha=3$, $\delta=1$. The new chain is therefore more specific 
than U(6) in identifying the lowest states and table 2 shows that it 
also contains a slightly greater percentage,  compared with U(6), in 
the numerical wave-functions  of  the  calculation \cite{EELL96}.  

    The position is less clear for non-yrast states.  The  zero entries 
for the $0^+_2$ states (in fact non-zero but very small) are largely a 
consequence of there being only one state with $\alpha=N$, $\delta=0$,
with most of it being taken up in the $0^+_1$ state. When we look at the  
content of the $0^+_2$ state, we find that, in \nuc{64}{Zn} and \nuc{66}{Ge}
it is well-described by a single O(18) $\supset$ O(15) label, about 98\%
$\alpha=N$, $\delta=3$, whereas in \nuc{62}{Zn} and \nuc{64}{Ge} there is
strong mixing between $\delta=3$ and $\delta=2$, both with $\alpha=N$.
For $2^+$ states, there are two independent states with the favoured O(18) 
$\supset$ O(15) labels of $\alpha=N$, $\delta=1$ as shown in table 1, and 
table 2 shows that the second state of this kind is generally the $2^+_3$. 
The nuclei with $T=0$ are exceptional, as in this case there is only one
state with the favoured labels, see ref.\cite{G96}. It is notable in table
2 that, in all five nuclei, these favoured states are concentrated more 
than 90\% in the lowest three calculated levels, whereas the total number
of 2$^+$ states in the model ranges from 7 in \nuc{62}{Zn} to 40 in 
\nuc{68}{Se}. The $2^+_2$ state is mainly described by $\alpha=N$, 
$\delta=2$ for which the percentages are 57, 89, 91, 95 and 85\% 
respectively for the nuclei in the table. As discussed earlier in this 
letter, the two states with $\alpha=N$, $\delta=1$ correspond roughly
to full U(6) symmetry $[N]$ and mixed symmetry $[N-1,1]$ and table 2
shows that the mixed symmetry lies mostly in 2$^+_3$ but with a larger
component in 2$^+_2$ for \nuc{62}{Zn}. In the two $T=0$ nuclei, both
$3^+$ states are extremely well described by $\alpha=N$, $\delta=3$.
This is possible because these states contain a large mixed symmetry
component, and it is clear from table 2 that mixed symmetry dominates 
these two states. For $T > 0$, there is an additional mixed symmetry 
state corresponding to $\alpha=N$, $\delta=2$ which, from table 2, accounts 
for 87\% of the $3^+_1$ in \nuc{62}{Zn}, while the $3^+_2$ is 97\%
$\alpha=N$, $\delta=3$. The $3^+$ states in both \nuc{64}{Zn} and 
\nuc{66}{Ge} are more strongly mixed between $\delta=2$ and $\delta=3$.

In this analysis of wave-functions from a numerical IBM-3 calculation,
based on a mapping from the shell model, the new O(18) $\supset$ O(15)
coupling scheme provides at least as good a first approximation, using   
largest O(18) label $\alpha=N$ and smallest O(15) label $\delta$, as the 
more familiar U(6) scheme. Assuming that this result is not a specific 
property of the rather simple shell model interaction used in 
ref.\cite{EELL96}, we conclude that it may be possible to use the
new group chain to simplify \mbox{IBM-3} calculations either through a
truncation of states or by reducing the number of essential parameters
in the hamiltonian. Some selection rules have already been given in 
ref.\cite{G96}. The essential physics of the O(18) $\supset$ O(15)
chain lies in the two-boson invariants, or pair states, for the two
groups, which are $(s^2-\sqrt{5}\, d^2)\,(J=T=0)$ and $d^2\, (J=T=0)$.
This contrasts with the more familiar O(6) and O(5) groups which,
in IBM-3, would also contain the corresponding pair states with 
$T=2$. Thus, the new chain discriminates between $T=0$ and $T=2$.
The new chain could be said to favour the $\gamma$-unstable O(6)
scheme for $T=0$ pairs and the U(5) scheme for $T=2$ pairs. It must
be remembered that, as for O(6), high seniority is lowest in energy
for O(18), so that the O(18) pair state is unfavoured. 

Finally, we comment that the general idea behind the new group chain 
\cite{G96} is not specific to IBM-3 and that a  corresponding new group 
chain O(12) $\supset$ O(10) exists in IBM-2 and would again imply mixing 
of U(6) which in IBM-2 implies mixing of F-spin. In fact, the corresponding 
table 1 for the IBM-2 example of $N_\nu =2$, $N_\pi = 1$ would be identical 
to that given here with the headings SU$_{\rm c}$(3) replaced by 
SU$_{\rm F}$(2), the F-spin group, O(18) replaced by O(12) and O(15) by O(10). 
In this case the new chain has pair states $(s^2-\sqrt{5}\, d^2)\,(J=0)$ and 
$d^2\, (J=0)$ for the $\nu$-$\pi$ system but not for $\nu$-$\nu$ or 
$\pi$-$\pi$, which is analogous with the IBM-3 argument given above.


\begin{thebibliography}{99}

\bibitem{IA87}
F. Iachello and A. Arima, {\it The Interacting Boson Model} (Cambridge 
University Press, Cambridge. 1987)

\bibitem{EW80}
J.P. Elliott and A.P. White, Phys. Lett. B97 (1980) 169

\bibitem{EES85}
J.A. Evans, J.P. Elliott and S. Szpikowski, Nucl. Phys. A435 (1985) 317

\bibitem{EELL96}
J.P. Elliott, J.A. Evans, V.S. Lac and G.L. Long, Nucl. Phys. A609 
(1996) 1

\bibitem{G96}
J.N. Ginocchio, Phys. Rev. Lett. 77 (1996) 28 

\end{thebibliography}
\end{document}